# An Efficient Method for Quantum Transport Calculations in Nanostructures using Full Band Structure

D. Basu, M. J. Gilbert, L. F. Register and S. K. Banerjee

Microelectronic Research Center, University of Texas at Austin, Austin, TX 78758 USA.

e-mail: dipanjan@mail.utexas.edu

**ABSTRACT**

Scaling of semiconductor devices has reached a stage where it has become absolutely imperative to consider the quantum mechanical aspects of transport in these ultra small devices. In these simulations, often one excludes a rigorous band structure treatment, since it poses a huge computational challenge. We have proposed here an efficient method for calculating full three-dimensionally coupled quantum transport in nanowire transistors including full band structure. We have shown the power of the method by simulating hole transport in p-type Ge nanowire transistors. The hole band structure obtained from our nearest neighbor $sp^3s^*$ tight binding Hamiltonian agrees well qualitatively with more complex and accurate calculations that take third nearest neighbors into account. The calculated I-V results show how shifting of the energy bands due to confinement can be accurately captured only in a full band full quantum simulation.

## I. INTRODUCTION

Electronic transport in mesoscopic systems has been an important field of solid state physics for the last two decades. Continuous scaling of semiconductor devices has resulted in the devices becoming so small that it has become necessary for device engineers and physicists to consider the effects of quantum transport in these mesoscopic systems. A theoretical understanding of the transport in the semiconductor devices being sought as improvement and/or augmentation of the Si CMOS technology, like ultrascaled Silicon (Si) FinFETs, Si and Germanium (Ge) nanowire transistors, is not complete without a full three dimensional quantum mechanical treatment of transport in these devices. Earlier, such quantum transport were modeled within the regime of effective mass [NEGF at EMA, Matthew JAP 04]. In spite of the approximations involved these approaches had proven helpful because for the device sizes being modeled, an approximate method like the effective mass approach gave a decent compromise between simplicity of the model and the reliability of results.

As devices continue to scale down, however, the use of effective masses in the transport calculations becomes more and more questionable. The proper electronic band structure in nanostructures can only be obtained by starting off with a model that describes accurately the full band structure for the bulk material, and then considering the effect of confinement on the energy dispersion relation. This can be obtained through a variety of approaches like the density functional theory, or empirical pseudopotential methods (EPM) or tight-binding (TB) methods. While *ab initio* density functional theories are now used to describe mesoscopic system without invoking empiricism, often these require prohibitive resources, and are limited by their inability to handle external electric biases and electric fields. On the other hand, the empirical atomistic methods, especially ones employing orthogonal tight binding basis allow a realistic and computationally efficient method for nanostructure simulations. However, even these atomistic models are limited by the complexity of the calculations, as realistic simulation of the nanostructures requires hundreds of atoms with their outermost valence orbitals, and the problem soon becomes intractable. Even with the advancement of the computational power of modern workstations, most of the nanostructure full band simulation procedures reported in the literature are run on highly parallel inter-communicating workstations or supercomputer facilities. These are too costly for regular device simulations, and with continuous shrinking of devices, we are in need of simulation procedures which will allow us to calculate full band full quantum transport of realistic nanostructures on powerful, yet personal workstations. In short, there is a need of more efficient ways of calculating full band quantum transport.

Most of the work published in this area treats quantum transport across the device as a single-particle ballistic transport situation. The electron in the device is injected from the two ideal wires or leads in the source drain regions (which are at different chemical potentials), and it undergoes scattering in the channel region from a spatially varying electrostatic potential. In the linear regime, one can express the conductance of the device G as a function of the total transmission probability (T) at the Fermi energy ($E_F$) using the Landauer formula:

$$G = \frac{2e^2}{h} T(E_F) \tag{1}$$

A common approach for the calculation of transmission is to use the Green's function for the scattering region and self-energies of the leads as [Caroli, Datta]:

$$T = Tr\left[\Gamma_L G^+ \Gamma_R G^-\right] \tag{2}$$

where $G$ is the Green's function of the channel region, and the elements of the matrix $\Gamma_{\{L,R\}}$ are given by:

$$\Gamma_{\{L,R\}} = i\left[\sum_{(L,R)}^+ - \sum_{(L,R)}^-\right] \tag{3}$$

and the self-energy terms $\sum_{(L,R)}^{+/-}$ are due to the semi-infinite leads on the left, L, and on the right, R respectively. The + (−) sign denotes the advanced (retarded) Green's functions and corresponding self-energies.

An alternative to the above procedure is to use a formulation by Ando [Ando 91], that is based on the matching of the wave function in the scattering region to the Bloch waves in the leads. The relationship of this mode-matching approach to the Green's function approaches is explained in detail in [Khomyakov, PRB 72, 0350450 (05)]. This technique has been successfully applied for conductance calculation using effective-mass tight-binding Hamiltonian [MacKinnon PRB 94, Matthew JAP 04, Usuki], as well on first-principles DFT model [DFT reference of Khomyakov, K Xia, PRB 06]. This method allows us to find the transmission across the scattering region, as well as the charge densities in the channel in real space through one single traversal of the device. In this paper, we extend this method by including full band structure using orthogonal nearest neighbor $sp^3s^*$ tight-binding orbitals as the basis set to express the wave-functions of the atomic orbitals in the device. While nearest neighbor $sp^3s^*$ basis is known to be insufficient for band structure description away from the $\Gamma$ point, this paper is aimed at showing the capability of the method in handling full band structure while retaining the complexity of the full 3-D quantum transport. Unlike other previous works [Nehari, SSE 06] we do not include the full band calculations only for the sake of extraction of appropriate effective masses for incorporating them in a simplified Hamiltonian later on. Instead the current and density respectively are calculated from the quantum mechanical current operator and the probability density associated with the atomic orbitals. Second nearest neighbor interactions that allows for a more proper rendition of the electronic band structure, and the inclusion of scattering will follow in a later work.

The paper is organized as follows. In Section II, we give the basic Hamiltonian of the system and show how it can be modified to form an eigenvalue system for the system in consideration. Section III deals with the criteria of selection of the eigenvalues in the contacts and how that procedure allows one to calculate the energy dispersion relation of these nanowires. The transmission matrix formalism for calculating transport across the device from source to drain is given in section IV. In section V, we show some I-V results for hole transport in p-type Ge nanowires computed with this method, and compare the characteristics with the results obtained for p-type Si nanowires. Finally we conclude in section VI.

## II. HAMILTONIAN

Following Ando's formalism [Ando 91], we break up the system into layers perpendicular to the transport direction. Naming each layer by the index $l$, we may write down the nearest neighbor tight-binding Schrödinger's equation for the $l$-th layer as:

$$H_{l,l-1}\psi_{l-1} + (H_{l,l} - E)\psi_l + H_{l,l+1}\psi_{l+1} = 0 \qquad (4)$$

Here the matrices $H_{l,l\pm1}$ denote the hopping elements of the Hamiltonian from layer $l$ to layer $l\pm1$, and $H_{l,l}$ denotes the onsite matrix elements of the Hamiltonian describing the system. $\psi_l$ denotes the wave-function of the atomic orbitals of the $l$-th layer, and $E$ is the injection energy (the Fermi energy $E_F$ for close to thermodynamic equilibrium situation). This description is fairly general, as a representation of any continuous Hamiltonian (e.g. one which we encounter in an effective mass approximation) that is discretised onto a real space grid lends itself to a tight-binding model. In such a case, $\psi_l$ represents the wavefunction of the lattice sites in the discrete real space grid. Alternatively, we will show here that the scheme can be perfectly used for a model that is discrete to start with, namely orbitals located on individual atoms in a three-dimensional solid state device.

The entire system is divided into three parts, a semi-infinite lead in the left and right, and a scattering region in the middle. Since nanostructure devices typically have a shape where there is a central constricted channel that flares out into wide source/drain region, our scattering region is not only the channel region, but also the source and drain regions. The left and the right leads are taken to be semi-infinite quantum wires that are in thermal equilibrium with externally applied bias, and are used for injecting carriers with an equilibrium distribution into the scattering region. A schematic of the device layers showing the finite scattering region sandwiched between the two long quantum wire leads is shown in Fig. 1.

Since the Hamiltonians are specific to the system under study, let us now concentrate on Ge nanowire as a specific device through which we want to exhibit the method. Ge has a face-centered cubic (fcc) lattice with a two atom basis. In this work we wish to calculate transport along [100] direction in Ge. Therefore each layer is composed of an fcc plane, and successive planes are displaced from the previous plane along the body diagonal by a distance equal in magnitude to $\left(a/4, a/4, a/4\right)$, $a$ being the lattice constant. Fig. 2 shows eight successive fcc planes along the transport direction that are stacked to form a square nanowire, which is 2 atoms wide. It is worth noting that for zinc-blend crystals like GaAs, successive layers are made up of anions or cations, while for monoatomic crystals like Ge, all planes are made up of identical atoms. It is evident that the atomic structure repeats itself every fourth layer. In the semi-infinite leads, where the potential variation along the transport direction can be taken to be constant (depending on the chemical potential), the wave-functions in each layer is related to the wave-function in the fourth preceding layer by the Bloch factor $\lambda$ (a constant phase difference), i.e., $\psi_l = \lambda \psi_{l-4}$.

Let us elaborate on the procedure for calculating the energy eigenstates in the leads. This is the first step for the transport simulation, as these wavefunctions define the transverse modes that are injected into the channel. Let us denote the four layers of the fcc lattice repeat unit by the numerals 1 to 4 (left to right), and the layer immediately preceding layer 1 as $\psi_{4,L}$ (layer no. 4 to the left), and the layer immediately following layer 4 as $\psi_{1,R}$ (Note that this has got nothing to do with the right lead, it just denotes the layer to the right of the four unit layers in consideration). Therefore, for the left lead, Eqn. 4 can be written explicitly as a system of four equations:

$$\begin{aligned}
H_{1,4}\psi_{4L} + (H_{1,1} - E)\psi_1 + H_{1,2}\psi_2 &= 0 \\
H_{2,1}\psi_1 + (H_{2,2} - E)\psi_2 + H_{2,3}\psi_3 &= 0 \\
H_{3,2}\psi_2 + (H_{3,3} - E)\psi_3 + H_{3,4}\psi_4 &= 0 \\
H_{4,3}\psi_3 + (H_{4,4} - E)\psi_4 + H_{4,1}\psi_{1R} &= 0
\end{aligned} \quad (5)$$

This system of four equations with six unknown wave-functions for the four layers of the primitive cell of the nanowire can be simplified using the following two Bloch-periodic relations:

$$\psi_{4L} = \frac{1}{\lambda}\psi_4, \quad \psi_{1R} = \lambda\psi_1 \quad (6)$$

Substituting $\psi_{4L}$ and $\psi_{1R}$ in Eqn. (5) gives us an eigensystem for the four layer wave-functions. However, we note from Eqn. (4) that wavefunctions of only two layers are necessary and sufficient for a complete description of the system. Algebraic manipulation allows us to reduce the eigensystem to a basis set consisting of the layer wavefunctions $\psi_2$ and $\psi_3$ as follows:

$$\begin{bmatrix} -H_{3,4}^{-1}H_{3,2} & -H_{3,4}^{-1}(H_{3,3}-E) \\ -H_{4,1}^{-1}(H_{4,4}-E)H_{3,4}^{-1}H_{3,2} & H_{4,1}^{-1}H_{4,3}-H_{4,1}^{-1}(H_{4,4}-E)H_{3,4}^{-1}(H_{3,3}-E) \end{bmatrix} \begin{pmatrix} \psi_2 \\ \psi_3 \end{pmatrix}$$
$$= \lambda \begin{bmatrix} H_{1,4}^{-1}(H_{1,1}-E)H_{2,1}^{-1}(H_{2,2}-E)-H_{14}^{-1}H_{12} & H_{1,4}^{-1}(H_{1,1}-E)H_{2,1}^{-1}H_{2,3} \\ -H_{2,1}^{-1}(H_{2,2}-E) & H_{2,1}^{-1}H_{2,3} \end{bmatrix} \begin{pmatrix} \psi_2 \\ \psi_3 \end{pmatrix} \quad (7)$$

We choose $\psi_2$ and $\psi_3$ as the bases to make the eigensystem in Eqn. (7) numerically well-balanced. In principle, any two successive layers could serve as the basis set.

### III. EIGENVALUES IN THE LEADS AND BAND STRUCTURE CALCULATION

Equation (7) is a generalized eigenvalue system whose dimensionality depends on the size of the system modeled, and the basis set used for modeling the individual atoms. For e.g., for a square Ge nanowire 3.4 nm wide, the no. of atoms in each fcc plane of Fig. 2 are $2\times 6^2 = 72$, and since we use 5 orbitals (sp$^3$s*) to denote the valence electrons (electrons in the outermost shell) in Ge, the dimensionality of the individual matrices of the Hamiltonian in this case is $n = 2\times 6^2 \times 5 = 360$. The generalized eigensystem of Eqn.(7) $(Ax = \lambda Bx)$ is of order $2n$, and we solve it using standard commercially available math libraries like IMSL [Visual Numerics ref]. Note that for this basis set, the onsite matrices $H_{l,l}$ are diagonal. The coupling matrices $H_{l,l\pm 1}$ are banded, and the distribution of non-zero elements in these sparse matrices depend on the geometry of simulation and the convention followed for numbering the individual atomic orbitals in the layers. The matrices of the generalized eigensystem are however not sparse.

The above eigensystem has $2n$ solutions of which $n$ are right going and $n$ left going (labeled as + and – respectively henceforth). These solutions can be further classified as propagating modes that are characterized by $|\lambda(\pm)|=1$, and evanescent modes otherwise. Acceptable right-going evanescent solutions are characterized $|\lambda(+)|<1$, since they decay to the right, and left-going evanescent wavefunctions have $|\lambda(-)|>1$. For distinguishing between the directions of the propagating modes, we refer to the probability current. The wavefunctions $\psi_l$ satisfy the time-dependent Schrödinger's equation:

$$i\hbar \frac{\partial \psi_l}{\partial t} = H\psi_l \qquad (8)$$

From this, we can derive the right-going probability current as:

$$j = \frac{1}{\hbar} \text{Im}\left[\psi_l^* H_{l,l+1} \psi_{l+1}\right] \qquad (9)$$

For our calculations, since $\psi_2$ and $\psi_3$ serve as the basis functions, we have $l=2$ in Eqn. (9). However, in principle, $j$ can be calculated between any two layers, since the probability current is conserved. The solutions to Eqn. (7) that have $|\lambda(\pm)|=1$, and $j>0$ are therefore right-going propagating waves. Equation (9) also serves us the purpose of normalizing the amplitude of the wavefunctions which is essential for calculation of carrier density in the system. The probability current in Eqn. (9) is, to be precise: the probability current carried per mode per unit energy by an occupied state, and this should be equal to $2|e|/h$, where $h$ is the Planck's constant. We adjust the coefficients of the propagating wavefunctions by a constant factor by calibrating the calculated probability current via Eqn. (9) to this constant factor $2|e|/h$.

Before we go to the transport calculation in the Ge nanowire MOSFET, it is interesting to see how the above calculation allows us to calculate the energy band dispersion for the nanowire, or any one-dimensional system in general. One can sweep the energy $E$ in Eqn. (7), and find the Bloch phase factors $(\lambda)$ of the modes allowed to propagate. If we take $x$ as the direction of transport for the nanowire, then the relationship of the wave vector $k_x$ to $\lambda$ is given by $\lambda = e^{ik_x \cdot a}$, since identical layers are $a$ distance apart. Thus the allowed values of $k_x$ when plotted against $E$, give the energy dispersion relation for a nanowire. In Fig.3 we show the energy dispersion relation for a (100) Ge nanowire using this method. The accuracy of the calculations is limited only by the accuracy of the nearest neighbor sp$^3$s* description of the bulk band-structure of Ge. The valence band effective masses are closer to the target [O. Madelung] than the conduction band effective mass values when using sp³s* basis set. This is because the energy bands which lie close to the Γ point are more accurately modeled than the energy extrema which lie away from the Γ point, e.g., close to the X point as in Si, or L point as in Ge. Therefore the valence band structure for the nanowire using our basis has good qualitative agreement with calculations involving up to third nearest neighbor interactions [Bescond Ge nanostructures]. Accuracy of the energy dispersion relation can be improved by a) including the five *d* orbitals at the cost of increase of the dimensionality of the matrices and eigenvalue system or b) including second nearest and/or higher neighbor interaction at the cost of increasing complexity of the problem without increasing the dimensionality significantly. We have started working on the latter approach and plan to report it in a later work.

### III. TRANSMISSION MATRIX FORMALISM

The eigenvectors in the lead can be classified as propagating and evanescent, and right and left-going (+ and − respectively) following the criteria given in Sec. II. For the sake of calculation of transmission, it is useful to write them in a form:

$$T_{l+0} = \begin{bmatrix} \psi_l(+) & \psi_l(-) \\ \psi_{l+1}(+) & \psi_{l+1}(-) \end{bmatrix} \qquad (10)$$

Here $\psi_l(+)$ is the $N \times N$ matrix that contains the modes moving to the right – the propagating as well as evanescent modes. Since our bases are the layer wavefunctions 2 and 3, $l=2$ in our calculations. It is interesting to note that while Eqn. (10) resembles the transmission matrix formalism used in literature[Matthew JAP 04, Usuki], unlike those, we do not require the wavefunctions in the successive layers, i.e., $\psi_l(+)$ and $\psi_l(-)$ to be related by the Bloch phase factor $\lambda$. Also, note that while in [Usuki] the matrix $T_0$ [see Eq. 2.11] is used for transforming from the mode space to real space; here we do the entire calculation in real space. We call the first matrix $T_{l+0}$ to distinguish it from $T_0$, and at the same time remind that it is composed of the basis wave functions corresponding to layer *l*. Having set up the basis set for the injection of carriers from the source lead, we can calculate the wavefunction in successive layers using:

$$\begin{bmatrix} \psi_l \\ \psi_{l+1} \end{bmatrix} = T_l \begin{bmatrix} \psi_{l-1} \\ \psi_l \end{bmatrix} \qquad 3<l<N+2 \tag{11}$$

where $l$ is the slice number in the device as we move from the source to the drain. The transfer matrix $T_l$ can be constructed from Eqn. (1) as:

$$T_l = \begin{bmatrix} 0 & 1 \\ -H_{l,l+1}^{-1} H_{l,l-1} & -H_{l,l+1}^{-1}(E - H_{l,l}) \end{bmatrix} \tag{12}$$

Therefore, using Eqn. (12), with $\begin{bmatrix} \psi_2 & \psi_3 \end{bmatrix}^T$, using $l=3$, we get $\psi_4$. Subsequently repeating the step with $l=4$ gives $\psi_5$ and so on. The net transmission from the source to the drain can be obtained by cascading the transfer matrices of the successive slices:

$$\begin{bmatrix} t \\ 0 \end{bmatrix} = T_{l+N+1}^{-1} T_{l+N} \ldots T_{l+1} T_{l+1} T_{l+0} \begin{bmatrix} 1 \\ r \end{bmatrix} \tag{13}$$

The current from source to drain is given by using Landauer formula for finite temperature as follows:

$$I_{SD} = \frac{2e}{h} \int dE\, T(E)[f_S(E) - f_D(E)] \tag{14}$$

The integration is done for energy range from the bottom of the band till where the Fermi functions ($f_S (f_D)$ are the Fermi functions at the source(drain) contacts) fall off to a very small value that can be approximated to zero, without introducing any error. The effect of broadening of the energy levels is accounted for in the limits of the integration.

Equation (13) is numerically unstable due to the presence of the evanescent modes that are present in the wavefunctions $\begin{bmatrix} \psi_l(\pm) & \psi_{l+1}(\pm) \end{bmatrix}^T$ which contribute to exponentially growing and decaying terms. Hence, instead of cascading the transmission matrices as in Eqn. (13), we follow the stabilization procedure followed in [Usuki, Matthew] where we set up the following iterative procedure [Usuki]:

$$\begin{bmatrix} C_1^{l+1} & C_2^{l+1} \\ 0 & 1 \end{bmatrix} = T_l \begin{bmatrix} C_1^l & C_2^l \\ 0 & 1 \end{bmatrix} P_l \qquad \text{for } 2 \leq T_l \leq N+3 \tag{15}$$

The final matrix $T_{N+3}$ in Eqn. (14) was used for going back to the mode space from the real (lattice) space [Usuki, Matthew]. In our calculation, we view it as the matrix that is used for convolving the wavefunction propagated from the source to the drain end with the wavefunction that exists in the right lead, i.e., the drain contact. This matrix is formed from the eigenfunctions at the drain lead and we calculate it using:

$$T_{N+3} = \begin{bmatrix} \psi_{N+3}(+) & \psi_{N+3}(-) \\ \psi_{N+4}(+) & \psi_{N+4}(-) \end{bmatrix}^{-1} \quad (16)$$

$P_l$ in Eqn. (15) is a linear operator that takes the following form to satisfy Eqn. (15):

$$P_l = \begin{bmatrix} 1 & 0 \\ P_{l1} & P_{l2} \end{bmatrix}$$

$$P_{l1} = -P_{l2} T_{l_{21}} C_1^l \quad (17)$$

$$P_{l2} = -\left[ T_{l_{21}} C_2^l + T_{l_{22}} \right]^{-1}$$

$T_{l_{21}}$ and $T_{l_{22}}$ are the components of the transfer matrices at each of the inner slices that are obtained from Eqn. (12) as:

$$T_l = \begin{bmatrix} T_{l_{11}} & T_{l_{12}} \\ T_{l_{21}} & T_{l_{22}} \end{bmatrix} \quad (18)$$

The iteration is started with the initial condition, $C_1^2 = 1$, and $C_2^2 = 0$ which corresponds incoming waves having amplitude unity. This iteration from $l=2$ to $l=N+3$ finally gives the transmission coefficient as $t = C_l^{N+4}$. A similar iteration for the reflection coefficient can be set up following [Usuki]

$$\begin{bmatrix} D_1^{l+1} & D_2^{l+1} \end{bmatrix} = \begin{bmatrix} D_1^l & D_2^l \end{bmatrix} P_l \quad \text{and} \quad r = D_1^{N+4} \quad (19)$$

Here, the initial condition is $D_1^2 = 0$ and $D_2^2 = 1$.

The transmission and reflection coefficients are calculated in the manner outlined above for an assumed potential profile in the channel, and the electronic charge density in the channel calculated from the transport calculation is fed back into the Poisson's equation to complete the self-consistent loop. For obtaining the density, instead of following Usuki's procedure, we employ an equivalent but computationally much more efficient way introduced in [Akis MSM 02]. Here for the final slice in the channel, we have [Akis]:

$$\psi_{N+2} = P_{N+2} \quad (20)$$

where $\psi_{N+2}$ is the matrix containing coefficients of the individual atomic orbitals of all the atoms in the layer $N+2$. Moving backwards from drain to the source, i.e., leftwards, one obtains [Akis]:

$$\psi_l = P_{l1} + P_{l2}\psi_{l+1} \tag{21}$$

The $P_l$ matrices being none other than the linear operators introduced in Eqn. (15) for stabilizing the calculations, one can store them in memory and not recalculate. Thus the probability density of the outermost shell electrons per atom in the plane is obtained by using:

$$n_{l,i} = \sum_{q,k=1,5} |\psi_{l,i,q,k}|^2 \tag{22}$$

Here, $k$ is used to denote the individual orbitals per atom (being sp$^3$s* basis, $k$ goes from 1 to 5), $q$ denotes the propagating modes, and $i$ denotes the numbering of the atoms in the individual fcc lattice plane of layer $l$. The total charge density $\rho$ that is fed into the Poisson's equation is obtained from $n_{l,i}$ by multiplying $n_{l,i}$ with the density of atoms per unit volume (which is roughly $8/a^3$, $a$ being the lattice constant).

## IV. RESULTS AND DISCUSSION

Figure 4 shows the schematic of the simulated square Ge nanowire transistor with gate all around the channel. All dimensions are given in A°. The source/drain regions are p+ doped to a concentration of $10^{20}$ cm$^{-3}$. The channel region is undoped. Explicit inclusion of the source/drain regions allows us to model constriction at the channel that can potentially cause carriers to transfer to different subbands while propagating. The source and drain regions are terminated by the semi-infinite quantum wire of square cross-section and width 39.55 A°. This can be viewed in the schematic on the right of Fig. 4. The channel region is wrapped on all four sides by a thin 5.65 A° gate oxide.

The $I_{SD}$-$V_G$ characteristics of the p-type Ge transistor outlined above is shown in Fig. 5. The device shows good subthreshold slope (~65 mV/decade), which can be attributed to the all-around-gate structure. The linear scale shows saturation characteristics for $V_{SG}$ above threshold voltage ($V_T$). A slightly high threshold voltage is to be expected as the bandgap for this 3.3 nm narrow Ge nanowire channel material is actually 1.05 eV instead of the bulk value of 0.66 eV. $V_{SD}$ for all of the simulations in this work was fixed at 0.25 V.

Within the transport calculation, we have simulated the confinement of the Ge channel by artificially raising the energies of the overlap integral of the bounding atoms, such that a band-gap typical of Ge and high-K dielectrics is created at the conduction and valence band edges (we selected conduction and valence band offsets to be 2 and 3 eV respectively, numbers that are approximate to high-K dielectric on Ge [Robertson]). We simulated a series of constricted channels that had the same oxide thickness; however, the channel thickness was reduced in order to see the effect of the full quantum model on the mode mixing that occurs when the transverse modes impinge from a wide source/drain region to a narrow constricted channel. These results are shown in Fig. 6, where we show the subthrehold characteristics of three devices that have the same 3.95 nm wide square nanowire as source drain regions, but channel widths varying from 3.39 nm to 2.26 nm. As expected, we find that the smaller channel carries lesser current because it has lesser cross-sectional area. This is also due to the fact that a narrower channel is more tightly confined, and hence a higher band gap material, and therefore, a higher gate bias is required to invert the channel and achieve same level of current as in the wider channel devices.

## IV. CONCLUSION

We have outlined here an efficient method to calculate fully quantum mechanical transport including full band structure by constructing a device atomistically. Unlike other full-band full quantum simulators, our method is fast and can be run on personal workstations. Though we have not done it, yet the method to calculate the transport is fully parallelizable as the transport calculation for injection at different energy values (for computing the current as per Eqn. 14 and the total carrier density) is independent of one another. Therefore device engineers can consider running on their personal workstations even a full band full quantum simulator for optimization of the nanowire transistors. We show the power of this method in treating hole transport across a p-type Ge nanowire transistor. Through one single probability current operator we are able to capture all forms of current, whether in the conduction or the valence band, or band-to-band tunneling. For (100) Ge nanowire transistors, we find excellent subthreshold characteristics, commensurate to the wrap-around gate structure. However, the calculations also show that as the devices are scaled down, confinement increases, increasing the spacing between energy bands, thereby increasing the threshold voltage of the devices. Therefore, for narrow channel devices, one requires a higher gate bias to invert the channel and have the same level of current as in a comparatively wider channel device.

LIST OF FIGURES:

Fig. 1. (Color Online) Schematic of the transport calculation showing how the device is broken up into a scattering region that includes source, channel and drain region, and semi-infinite source and drain contacts at the two ends. The H's denote the onsite Hamiltonian corresponding to the four different layers for the four fcc lattice plane that make the repeat unit in these nanowires.

Fig. 2. Eight fcc lattice planes, along (100) direction, showing the individual atomic locations in the plane. For a diamond crystal structure all the atoms are identical. For GaAs say, the layers are alternately anions and cations.

Fig. 3. Energy dispersion relationship for the valence band of a square (100) Ge nanowire of width 3.96 nm.

Fig. 4. Schematic of the device simulated showing in left: the cross-section of the channel region, perpendicular to the transport direction, and in right, the view from the top showing the source and drain contact at the ends, and the gate oxide in the middle, surrounding the channel. All dimensions are in $\text{A}°$.

Fig. 5. $I_{SD}$-$V_G$ in linear and log scale for a p-type (100) Ge nanowire of square cross-section. Width of the channel is 3.39 nm. $V_{SD} = 0.25$ V.

Fig. 6. $I_{SD}$-$V_G$ for three p-type (100) Ge nanowire transistors. The devices all have square cross-sectional area with source/drain regions 3.96 nm wide, and the channel regions constricted to the widths shown in the legend. $V_{SD} = 0.25$ V as before.

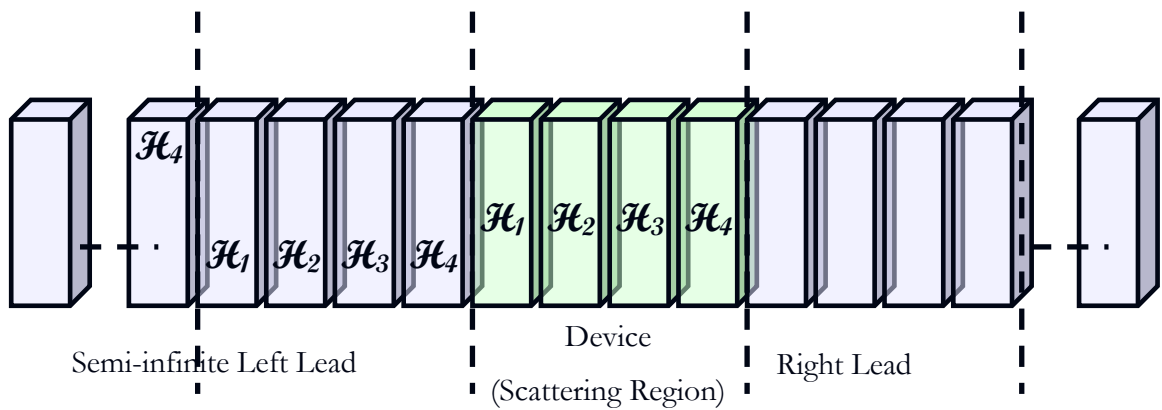

Figure 1.

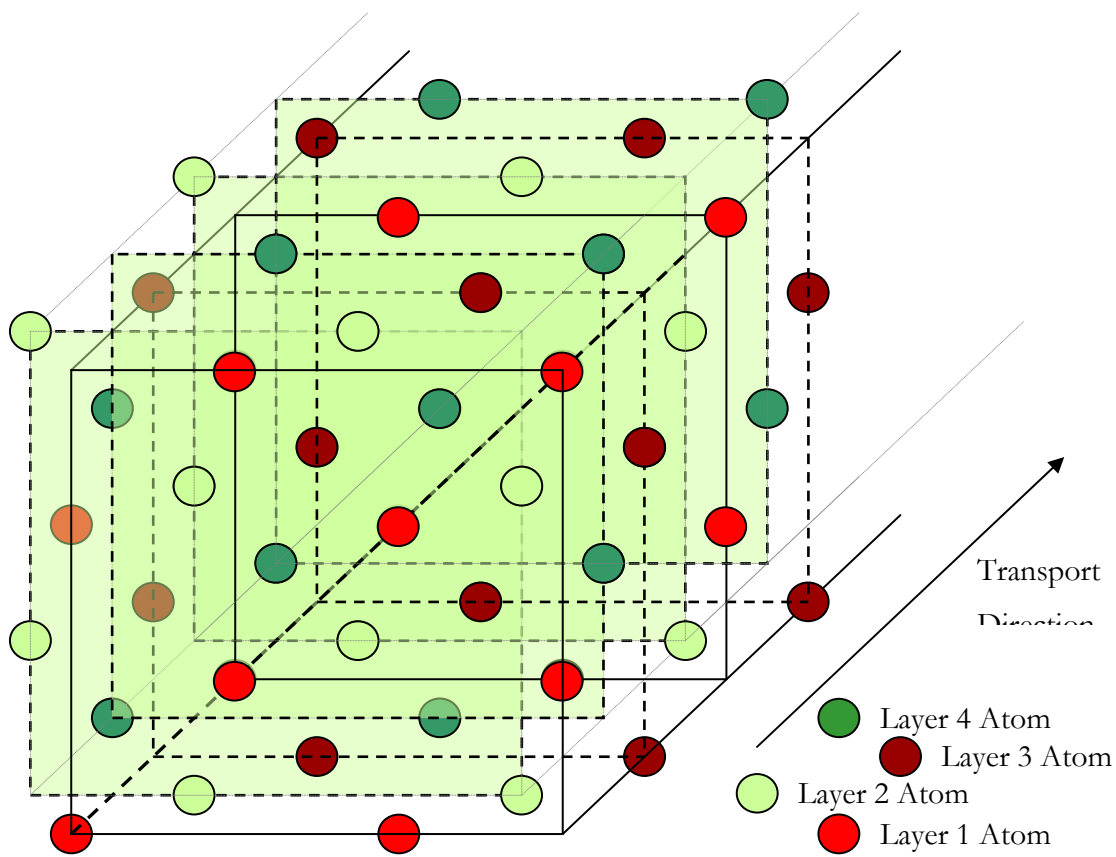

Figure 2.

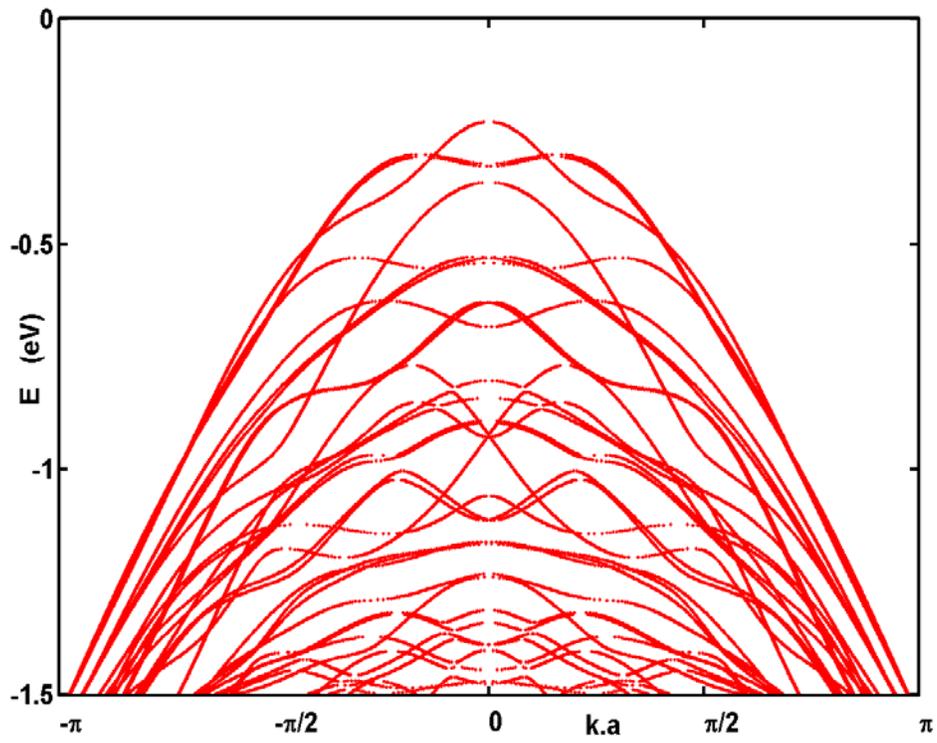

Figure 3.

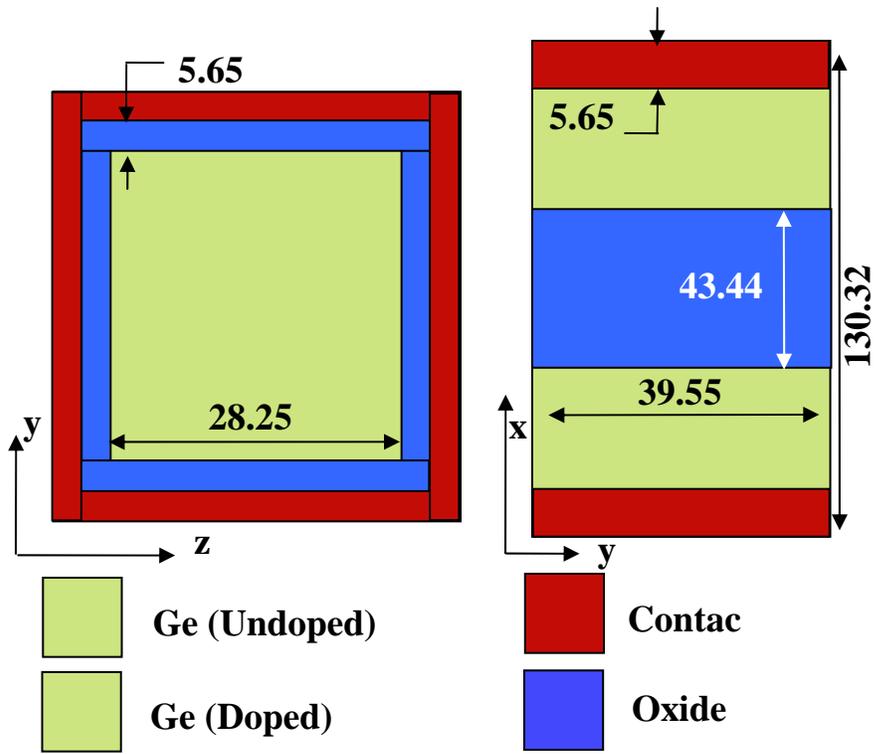

Figure 4.

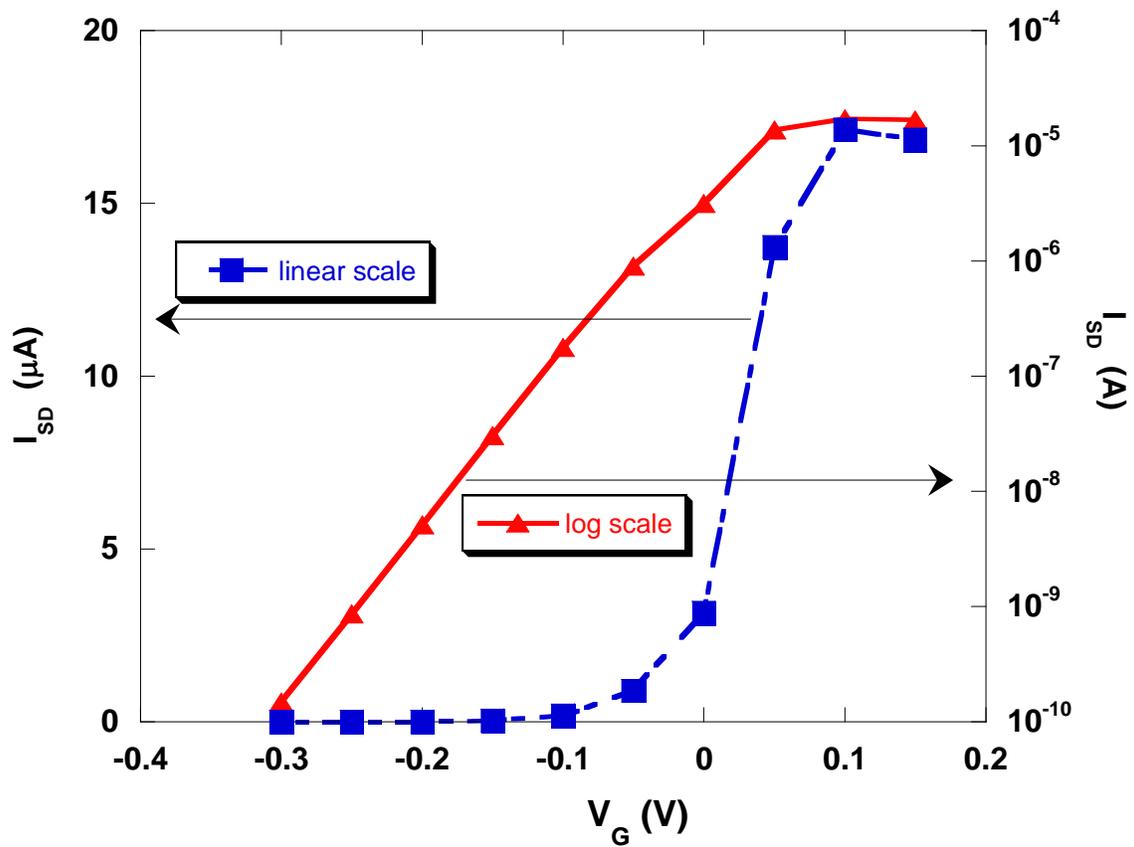

Figure 5.

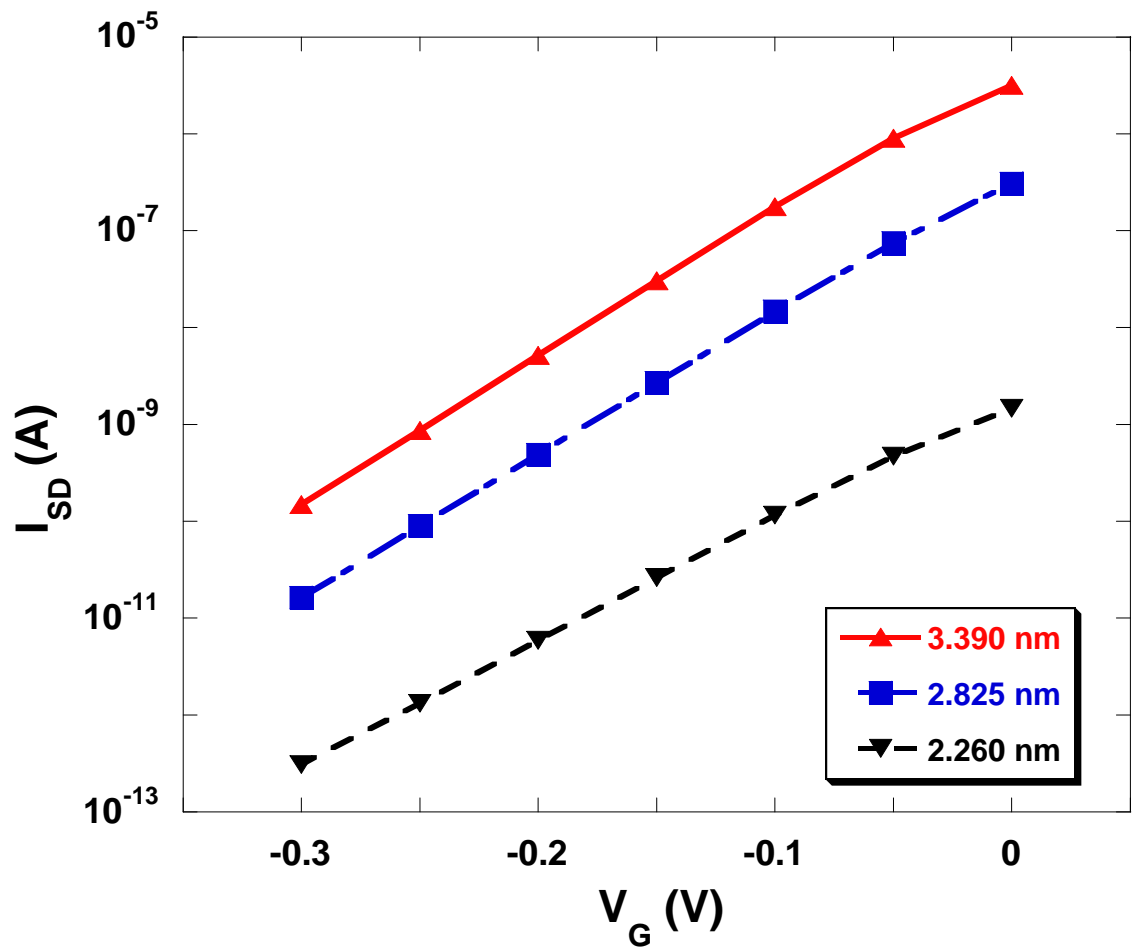

Figure 6.